\pdfoutput=1

\documentclass[11pt]{article}

\usepackage[final]{acl}
\usepackage{multirow}

\usepackage{times}
\usepackage{latexsym}
\usepackage{amsmath}
\usepackage{booktabs}
\usepackage{float}
\usepackage[T1]{fontenc}

\usepackage[utf8]{inputenc}

\usepackage{microtype}

\usepackage{inconsolata}

\usepackage{graphicx}
\usepackage{hyperref}
\usepackage{listings}
\usepackage{xcolor}  
\usepackage{multirow} 
\usepackage{fontawesome}
\usepackage[ruled,vlined]{algorithm2e} 
\usepackage{amssymb}

\title{Tree-of-Code: A Self-Growing Tree Framework for End-to-End Code Generation and Execution in Complex Tasks}

\author{
 \textbf{Ziyi Ni\textsuperscript{1,2,}}\thanks{These authors contributed equally to this work.}\thanks{This work was conducted at Qianfan AppBuilder Group during the Baidu AI Cloud Summer Camp internship.},
 \textbf{Yifan Li\textsuperscript{4,}}\footnotemark[1]\footnotemark[2],
 \textbf{Ning Yang\textsuperscript{1}},
 \textbf{Dou Shen\textsuperscript{3}},
  \textbf{Pin Lv\textsuperscript{1,‡}},
 \textbf{Daxiang Dong\textsuperscript{3,‡}},
\\
 \textsuperscript{1}The Key Laboratory of Cognition and Decision Intelligence for Complex Systems, \\Institute of Automation, Chinese Academy of Sciences
 \\
  \textsuperscript{2}School of Artificial Intelligence, University of Chinese Academy of Science \\
 \textsuperscript{3}Baidu, Inc.
 \textsuperscript{4}Global Innovation Exchange Institution, Tsinghua University
\\
 \small{
   \textbf{‡ Correspondence: }\href{mailto:email@domain}{dongdaxiang@baidu.com}, \href{mailto:email@domain}{pin.lv@ia.ac.cn} 
 }
}

\begin{document}
\maketitle
\begin{abstract}
Solving complex reasoning tasks is a key real-world application of agents. 
Thanks to the pretraining of Large Language Models (LLMs) on code data, recent approaches like CodeAct successfully use code as LLM agents' action, achieving good results. 
However, CodeAct greedily generates the next action's code block by relying on fragmented thoughts, resulting in inconsistency and accumulative hallucination.
Moreover, CodeAct lacks action-related ground-truth (GT), making its supervision signals and termination conditions questionable in multi-turn interactions.
%
To address these issues, we propose Tree-of-Code (ToC), a self-growing framework that generates nodes through self-supervision, incorporating prompt and model exploration in a GT-free setting. Each node employs CodeProgram, an end-to-end code generation paradigm that aligns executable code logic with global reasoning. This approach uses task-level execution success as both node validity and stop-growing flags, bypassing process supervision to enable online applications.
Experiments on two datasets with ten popular zero-shot LLMs show that ToC boosts accuracy by nearly 20\% over CodeAct with fewer than 1/4 turns. To further investigate the trade-off between efficacy and efficiency, ablation studies on different ToC tree sizes and exploration mechanisms validate ToC's superiority. 

\end{abstract}
\section{Introduction}

Large language models (LLMs) significantly improve agents' ability to leverage external tools. \cite{chen2023autoagents,hong2023metagpt,paul2024continually}.
Effectively and efficiently handling complex real-world problems \cite{blount1994agi}, especially those requiring multiple tools and calls \cite{li2023api, wang2024codeact}, has become a key focus across industry and academia.
Currently, the widely used paradigm, ReAct, \cite{yao2022react}, combines reasoning with action strategies, allowing for actions to be performed incrementally and adjusted based on environmental feedback. 

The application of code generation techniques to complex task planning and execution has garnered significant attention~\cite{holtl2mac,wen2024learning,xu2024wizardlm}, particularly with the emergence of CodeAct~\cite{wang2024codeact} approaches. 
CodeAct moves the interaction unit from ReAct's individual tool calls to generating code blocks with local reasoning while leveraging code logic and libraries.
Rather than JSON \cite{Qin2023ToolLLMFL} or text \cite{Park2023GenerativeAI}, it treats code as action, utilizing LLM's pre-trained coding skills for efficient handling of complex tasks.

However, CodeAct treats each turn as an individual action rather than addressing the entire program, following a step-by-step generation process. While this approach may seem explorative, it has four critical limitations: 
(I). CodeAct assumes that the ground truth (GT) is known and uses GT matching as a termination criterion, which is unrealistic and unfeasible.
(II). Fragmented thinking is inefficient. For simple problems, stalled thinking is not only unnecessary but also disrupts the logical chains in the code \cite{wang2023leti, guo2024deepseek}. 
Moreover, as the number of turns increases, repeatedly integrating prior thoughts causes context overload, heightening model hallucinations \cite{ji2023survey}, increasing computational cost.
(III). CodeAct lacks exploration of diverse reasoning paths. While it supports multi-turn interactions, it follows a single reasoning process. In contrast, solving complex problems often has multiple solutions \cite{Mialon2023GAIAAB}, where different approaches can branch from different points, making it difficult to set a standard answer for each turn. 
(IV). Generated trajectory data is hard to reuse. When using these trajectories for supervised fine-tuning (SFT), they cannot be directly combined into a single program response \cite{wang2024codeact}. Reinforcement learning is also challenging due to the lack of process supervision \cite{zelikman2024quiet}, leading to fundamental issues.

Since defining and obtaining supervision signals for intermediate states is challenging, we propose using task-level feedback directly, treating task completion as a single step. We introduce CodeProgram, an end-to-end code reasoning and generation paradigm, as a ‘turn,’ where the only environmental supervision is execution success. To incorporate reflection and exploration, we design an outcome-driven refinement framework, Tree-of-Code (ToC), that enables multi-turn interactions with diverse solutions exploring the model and prompt pools as tree branches, where task-level CodePrograms serve as the nodes. The final output is determined by voting on the collected nodes, selected based on their successful execution.
It’s important to note that, in this paper, a ‘turn’ refers to a single action of code generation. For our CodeProgram, a turn involves completing the entire program, rather than just a single task step (as in CodeAct).

Although ToC’s name and structure are similar to "Tree-of-Thoughts" (ToT) \cite{yao2024tree}, their meanings fundamentally differ. Our concept might be closer to a Code "Random Forest" \cite{rigatti2017randomforest}.
While ToT enhances "Chain-of-Thought" (CoT) \cite{wei2022cot} by exploring different thoughts within the same solution, ToC explores multiple distinct program solutions. In other words, each node in ToC represents a complete solution, and the tree as a whole captures different iterative optimizations (depth) across a variety of complete solutions (breadth). 
The core contributions of this paper are summarized as follows: 
\begin{enumerate}
\item We propose a self-growing Tree-of-Code (ToC) structure that automatically reflects and explores diverse, complete solution nodes without labeled data, facilitating complex tasks in multi-tool online scenarios. 
\item Each node in ToC, called a CodeProgram, is generated end-to-end. We are the first to define process-level supervision at the task-outcome level using execution success.
\item Extensive experiments and ablation studies on two multi-tool, complex task datasets with ten models, demonstrate that ToC significantly enhances problem-solving accuracy and efficiency in real-world, zero-shot scenarios.
\end{enumerate}

\section{Design Motivation}

\begin{figure}[t]
  \centering 
  \includegraphics[width=0.48\textwidth]{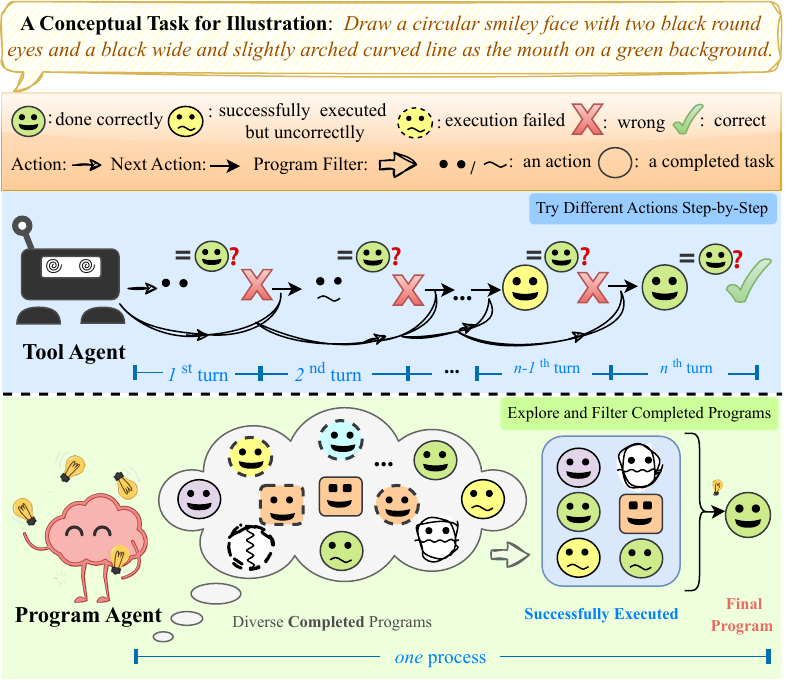}
  \caption{Illustration of our design motivation.
  }
  \label{fig:cute}
\end{figure}

\begin{figure*}[ht]
  \centering
  \includegraphics[width=1\textwidth]{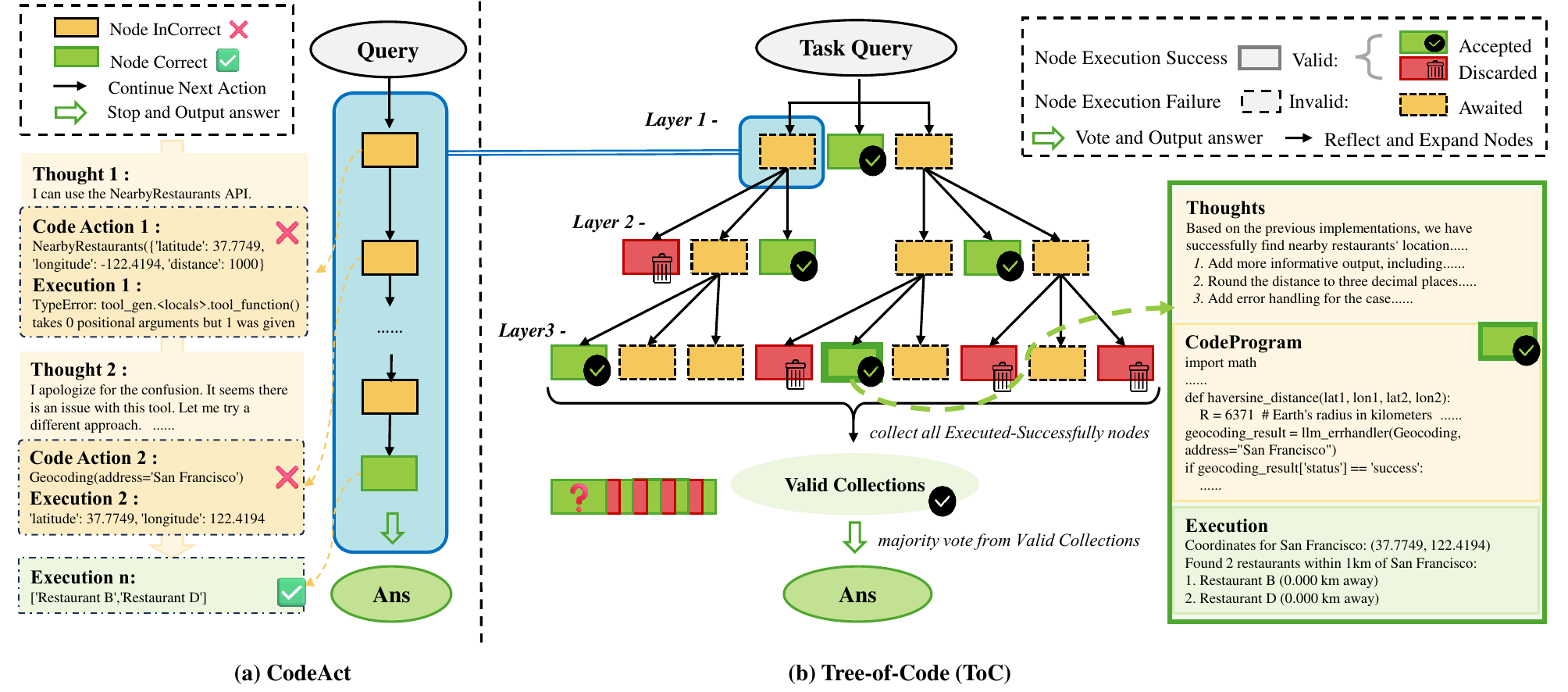}
  \caption{An Overview of \textbf{CodeAct} and \textbf{ToC}.
(a) CodeAct regards code as action with step-by-step reasoning.
(b) ToC applies execution-based reflection in the tree structure, where each node (CodeProgram) generates end-to-end code with global planning as its thoughts. 
At each layer, nodes are executed in parallel; if executed successfully, they are collected for voting. 
Note that the process supervision relies solely on the node's execution success or failure, rather than on the specific content executed (whether correct or incorrect), which would require pre-known labels.  
The query is "Find nearby restaurants within 1km of San Francisco" from API-Bank level-3 dataset. } 
\label{fig:overview}
\end{figure*}

In industry, complex tasks requiring multiple tools and function calls, are typically driven by open-ended user queries. This creates two key challenges: (1) 
For zero-shot queries, it is unrealistic to pre-obtain task-level ground-truth (GT), which is required for SFT \cite{chung2024flant5} or reinforced fine-tuning (ReFT) \cite{Luong2024ReFT}. Moreover, without GT, the termination criteria become unclear.
(2) 
Multi-turn interactions lack a standard trajectory, making it difficult to define the process supervised signals \cite{Luo2024ImproveMR}. 
Current methods often rely on 'LLM-as-judge' to evaluate whether the user’s needs are met at each step \cite{Chen2024JumpCoderGB, Li2024CodeTree}. 
However, it would require an API call after every step to check progress, ultimately increasing both time and token costs. Besides, the evaluation without objective signals demands strong analytical and reasoning skills from LLMs.
Existing methods deliberately avoid these challenges by assuming GT is known~\cite{wang2024codeact}, matching task-level GT with action-related outcomes at each step, like the tool agent in Figure \ref{fig:cute}: the interaction turn stops only if they match, or continues until the step limit is reached. 

Since intermediate states are absent, if possible, why not treat each complete end-to-end execution as an atomic state? By iteratively exploring feasible solutions through parallel executions, we first collect a batch of solutions, and then determine the optimal one, as shown by the program agent in Figure \ref{fig:cute}. This idea inspires our node outcome-driven reflection system specifically designed for multi-tool interaction in real-world environments. Our key contribution is a self-growing framework enabling LLM agents to autonomously interact with code through zero-shot learning without GT supervision, whose implementation details will be subsequently presented.
\section{Tree-of-Code Method}
Following the design motivation, we need to collect all valid solutions and identify the one closest to the GT. 
By treating each tree node as a complete task-level solution and exploring different nodes for breadth while deepening through iterative refinement, we propose ToC (Tree-of-Code), an execution-based, self-growing, and self-filtering tree for handling real-world complex tasks. 

\begin{figure*}[th]
\centering
  \includegraphics[width=0.98\textwidth]{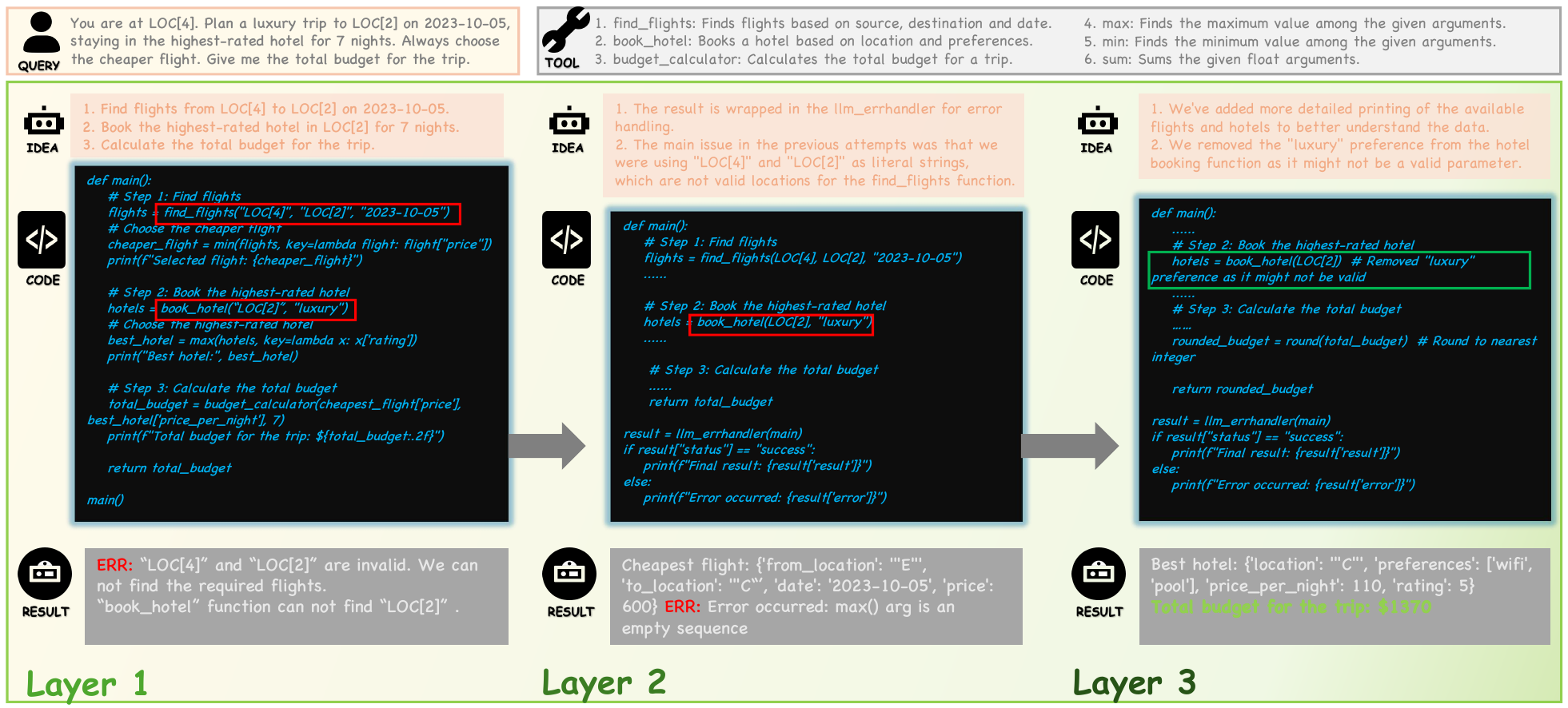}
  \caption{Illustrative example of a branch of ToC. We demonstrated the process of a node expanding into deeper levels. Based on the user query, tool descriptions, and previous execution outcomes, ToC first thinks about how to do it and then writes the end-to-end code. The example is selected from M3ToolEval dataset.
  }
  \label{fig:example}
\end{figure*}

\subsection{Overview of Tree-of-Code}
We represent the ToC framework as $\textbf{\textit{T}}=(\textbf{\textit{N}, \textit{S}})$, where $\textbf{\textit{N}}$ denotes a set of nodes ($N$), and $\textbf{S}$ represents the stems (unidirectional arrows in Figure \ref{fig:overview})
,  modeling the reflection reasoning process of LLMs when expanding the nodes. 
The overview of ToC and how it works is illustrated in Figure \ref{fig:overview}. 
Let $L$ denote the max depth,
$l$ the layer index, $M$ the expanded layer's max-width, 
$m$ the node index,
$l \in \{1, \dots, L\}$, $m \in \{1, \dots, M\}$. 
We use 
$T$ for the thoughts of the $N$, 
$C$ for code, and 
$E$ for its execution result.  
The next-layer $N$ is denoted as: 
\[N_{(l+1)\text{-}m} =S_{l\to {(l+1)}} (f, \sum_{j=0}^{l} (T_{j\text{-}m} + C_{j\text{-}m} + E_{j\text{-}m}))\]
where $f$ represents the basic information of the task, such as the user's query,  and all tool descriptions. The sum $\sum_{j=0}^{l}$ indicates that each reflection reasoning process for generating the next node relies on the thoughts, code, and execution results from all ancestor nodes in the history. 
The node index is fixed for simplicity in the formula.

\subsection{Tree Node Generation}
\label{codeprogram}


Unlike tool agents like CodeAct, which treat each intermediate action and environmental feedback as a step, each node in our ToC represents a complete task, effectively increasing the granularity of task handling at each layer. 

\begin{figure}[t]
  \centering  \includegraphics[width=0.48\textwidth]{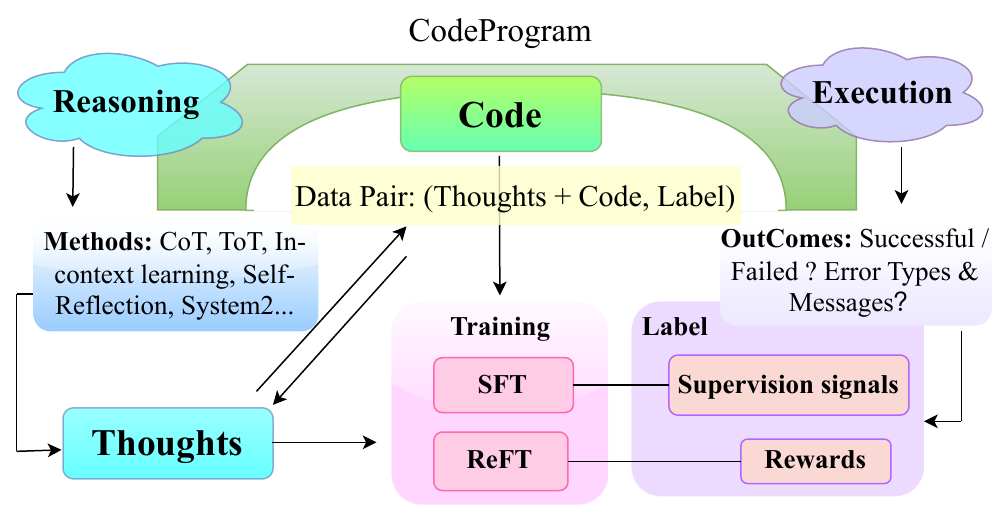}
  \caption{Illustration of the CodeProgram in ToC.}
  \label{fig:codeprogram}
\end{figure}

In other words, a single tree node (one turn) is equivalent to multiple turns of CodeAct, with both being directly comparable and serving the same purpose (final response), significantly improving efficiency. We refer to this end-to-end code reasoning and generation paradigm as CodeProgram. 
Figure \ref{fig:codeprogram} illustrates how it works.

Specifically, the end-to-end code in CodeProgram serves as a bridge, aligning with natural language reasoning and execution outcomes in the environment. 
Besides, by decoupling the reasoning process from code execution, 
we achieve flexibility while ensuring consistency. 

\subsubsection{Code as Reasoning}

On the one hand, CodeProgram leverages the concept of "code-as-reasoning" to generate code, where the process of writing code itself mirrors the reasoning process.

On the other hand, global reasoning is essential for guiding CodeProgram's complete code generation in a single end-to-end flow. This approach enables the seamless integration of various reasoning techniques for large language models (LLMs), such as prompt engineering \cite{chen2023prompteng}, Chain-of-Thoughts (CoT) \cite{wei2022cot}, Tree-of-Thoughts (ToT) \cite{yao2024tree}, in-context learning \cite{kojima2022step_by_step}, self-reflection \cite{zhang2024selfcontrastreflect}, and System2 reasoning \cite{frankish2010system2reason, openai2024o1}. Additionally, longer chains of thought have consistently been shown to enhance task performance \cite{zelikman2024quiet}.

Building on this foundation of global reasoning, we write the root prompt based on previous work \cite{wang2024codeact} to guide the generation of step-by-step CoT thoughts and the corresponding complete code. 
LLMs are prompted to first analyze and break down the problem, generate reasoning-based thoughts for solving it, and then produce the complete code that reflects and executes that reasoning.
The thoughts and codes are enclosed using the "\textit{<thought>-</thought>}" and "\textit{<execute>-</execute>}" tags, respectively.
The root prompt is shown in Appendix  \ref{Appendix Prompt}.

\subsubsection{Two Helper Tools}
\label{Two Helper Tools}
CodeProgram struggles with environmental exploration when LLMs must rely on tool outputs to determine the next steps. For instance, in web browsing tasks, the next action can only be decided after viewing the page content, and a final summary answer can only be provided after considering all tool outputs. 
Thus, to maintain end-to-end flow, we introduce two functions: a general \textbf{res\_handler}, which defines a prompt to generate results that meet the prompt requirements for final summarization, and a specific \textbf{next\_action} for web tasks, 
which decides the next action from a given set of possible browsing actions based on the page content, visited URLs, and task query. 
Their tool descriptions and functions are shown in  Appendix \ref{Helper Tool}. 

They help better understand the semantic relationships between tools, ensuring a smooth, cohesive sequence of tool calls during code generation. In the Appendix \ref{Helper Tool Example}, we also provide an example demonstrating how these helper tools work.

\subsubsection{Execution Outcome as Process Label}
The code solution is task-level, and its execution outcome is a self-provided annotation that can be directly used as labels. 
Note that we focus solely on task execution success, using a simple true/false label to filter feasible solutions and approximate more effective ones. This label is weak but available, simple, and useful—unlike pre-known GT or correctness judgments. 

Benefiting from our end-to-end paradigm (a direct, complete task-level response to a single query), we can select "successfully executed" samples for SFT and use various rich comments (such as specific results or error messages) as rewards for ReFT by repeating the CodeProgram in different settings (i.e., multi-nodes). In this context, the code acts as a verifier. This verification-then-refinement concept also inspires the development of a multi-layer Tree-of-Code (ToC). 



Thanks to task-level granularity, the code's execution outcomes align with both the task query and the thought-code output, enabling the generation of valuable data for potential future training. 


\subsection{Tree Expansion}
We initialize from a root node and recursively expand the tree. 
The expansion process follows:
(1) The breadth-first search (BFS) strategy is applied, with each parent node branching into $M = 3$ child nodes.
(2) Whether the node continues to grow depends solely on the evaluation of its own execution state (success or failure). 
For each $N_{l}$,   
\[
\left\{
  \begin{array}{ll}
    \text{\textit{stop and collect},}
    & \text{\textit{if} } E_{l} \neq \text{\textit{None or error}},\\
  \text{\textit{grow} } \textbf{\textit{N}}_{(l+1)}, & \text{\textit{otherwise}.}
  \end{array}
\right.
\]
(3) Expansion continues until all child nodes stop or the maximum depth ($L$) of 3 is reached.


\begin{figure}[ht]
  \centering  
  \includegraphics[width=0.40\textwidth]{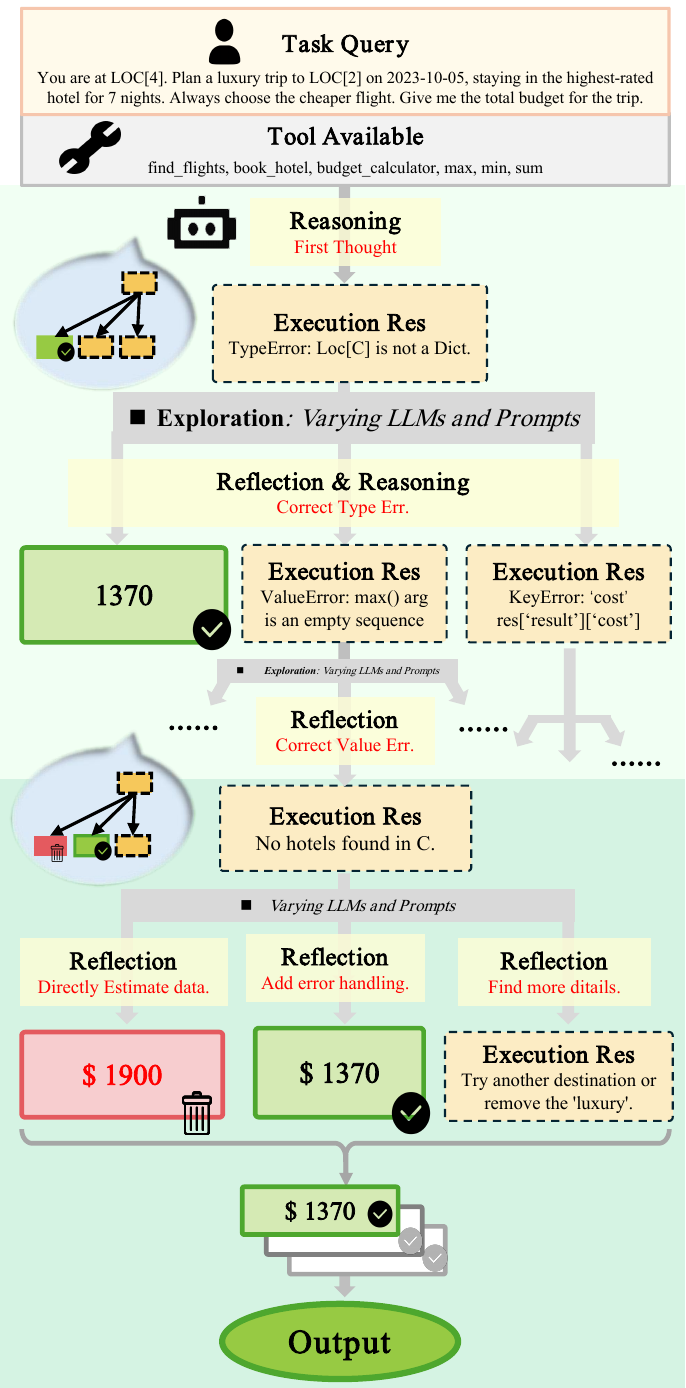}
  \caption{A detailed example illustrating ToC's execution-based reflection and expansion. 
  }
  \label{fig:tree}
    \vspace{-2pt}
\end{figure}

\vspace{3pt}
\noindent \textbf{Execution-based Reflection.} 
We can not guarantee that one node solution will be correct on the first attempt. 
Treating task-level execution errors as continuation signals,
we propose execution-based reflection, which enables LLMs to self-reflect, identify errors, refine thoughts, and improve code. As long as execution fails, self-reflection continues iteratively, generating next-layer new nodes. 
The prompt for reflection is shown in Appendix \ref{Reflection Prompt}.

This also allows the branch to grow into deeper layers, where each node in the trajectory provides process supervision signals based on its outcome. Note that the definition of 'turn' is equivalent to that of 'layer'; both terms carry the same meaning. 
Since these supervision signals are inherently embedded within the CodeProgram node, the growth process is self-driven. Therefore, the whole tree is end-to-end generated. 

Figure \ref{fig:example} shows an example of a branch of ToC while Figure \ref{fig:tree} demonstrates execution-based reflection and tree expansion. 
Additionally, our flexible tree-structured framework allows for the integration of any reflection method for tree expansion.

\vspace{2pt}
\noindent \textbf{Exploration strategy.} 
Generating code in a single pass presents two main limitations on diversity: 

\begin{itemize}
    \item 1) Limited strategy: It easily leads to cognitive narrowing, where the fundamental reasoning mechanism remains unchanged.
    \vspace{-2pt}
    \item 2) Limited robustness: 
    If an error occurs, the only option for the user is to re-run the whole process, without any proactive adjustments, which leads to inefficiencies. 
\end{itemize}

Research \cite{renze2024self} has shown that performance benefits from diverse perspectives of error identification, which encourages models to generate multiple solutions (ie. nodes in ToC). 

To enhance the diversity of ToC, we introduce randomness into the expansion process by varying LLMs and prompts, inspired by the random forest \cite{rigatti2017randomforest}. 
At the system level, different LLMs from our list, introduced in Section \ref{section models}, are explored randomly with a consistent temperature setting of 0.1. At the instruction level, prompts are randomly selected from a diverse pool, created through self-evolution and human crafting. 

The random exploration mechanisms operate at each node individually, while the prompt pool is created just once for the entire system. 

Specifically, we used ten LLMs to generate ten diverse prompts through prompt evolution from the root prompt (see Appendix \ref{Appendix Prompt}). The evolution process ensures the core content remains consistent while promoting orthogonal or divergent expressions. 
Six distinct prompts were manually selected and the following modifications were then applied: (1) adding detailed usage examples (beyond just printing "Hello world") to three prompts; (2) adjusting the format with line breaks and indentation; (3) randomly rearranging components, including the reflection part, usage examples, role instructions, tool descriptions, and chat history.

\subsection{Final Result Generator}
Once valid outputs from successfully executed nodes are collected, the same LLM makes the final decision by performing a majority vote and summarization to determine the most likely answer. Ties are rare in our observations, so we always choose the most frequent answer without special handling. 

\section{Experiment and Analysis}
\subsection{Setup}
\begin{figure}[ht]
\centering
  \includegraphics[width=0.45\textwidth]{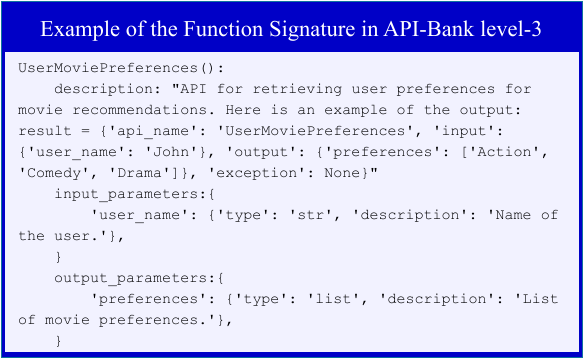}
  \caption{Example of the function signature in level-3.
  }
  \label{fig:function}
\end{figure}

\textbf{Datasets. } 
Following CodeAct, our evaluation is based on M3ToolEval\footnote{\url{https://github.com/xingyaoww/code-act/tree/main/scripts/eval/m3tooleval}}  (M3) 
\cite{wang2024codeact}  
and the test set of API-Bank\footnote{\url{https://huggingface.co/datasets/liminghao1630/API-Bank/tree/main}
} \cite{li2023api}.
M3 consists of 82 tasks utilizing 100 tools in code/JSON/txt action space respectively across 5 types of scenarios, including DNA sequencer, message decoder, trade calculator, travel itinerary planning, and web browsing.  
API-Bank contains 314 tool-use dialogues and 73 API tools, including level-1, 2, 3.
Unlike CodeAct, which evaluates only on level-1, we focus directly on the 50 most challenging level-3 tasks, 
on which nearly all non-GPT4 models score 0\%, according to the original paper.
Considering API-Bank only supports JSON format, 
we make following modifications to adapt it for code interaction:
(1) functionalize all API tools, 
(2) add output examples to each function description (Figure \ref{fig:function}). 
We include all tool signatures in the prompt context and let LLMs inherently search and select tools, instead of using ToolSearch API, deemed the least essential in \cite{li2023api}.
(3) determine correctness by matching the response to the expected final output through conditional keywords, not by API call matching.

\begin{table*}[ht]
  \centering
  \resizebox{0.85\textwidth}{!}{  
    \renewcommand{\arraystretch}{1.0}
    \begin{tabular}{lccc|ccc}
      \toprule
      \multirow{2}{*}{\textbf{Mechanism}} & \multicolumn{3}{c}{\textbf{M3ToolEval}} & \multicolumn{3}{c}{\textbf{API-Bank level-3}} \\
      \cmidrule(r){2-4} \cmidrule(r){5-7}
      & \textbf{Avg Turns} & \textbf{Correct} & \textbf{Output Words} & \textbf{Avg Turns} & \textbf{Correct} & \textbf{Output Words}\\
      \midrule
      \textbf{ReAct} & 8.2 & 38.1 \% & 1.86 k & 9.5 & 8.2 \% & 1.66 k \\
      \textbf{CodeAct} & 7.0 & 49.4 \% & 1.91 k & 8.9 & 19.2 \% & 1.82 k \\
     \textbf{Tree-of-Code (3-3)} & 1.7 \textbf{\(\downarrow\)} & 67.1 \% \textbf{\(\uparrow\)} & 0.44 k \textbf{\(\downarrow\)} & 2.1 \textbf{\(\downarrow\)} & 38.0 \% \textbf{\(\uparrow\)} & 0.39 k \textbf{\(\downarrow\)} \\
      \bottomrule
    \end{tabular}
  }
  \caption{
    Performance comparison of the baselines and our ToC in terms of averaged turns, output words, and accuracy on two datasets. Note: all numerical results presented in this paper are rounded. 
  }
  \label{tab:main}
\end{table*}

\noindent \textbf{Models. }
\label{section models}
We include the following ten models in our model pool for evaluation:
the GPT family from OpenAI \cite{achiam2023gpt, bubeck2023sparks, openai2024hello}, including gpt-3.5-turbo-1106, gpt-4o-mini-2024-07-18, gpt-4o-2024-08-06, and gpt-4-1106-preview checkpoints, excels in generation capabilities. From the Anthropic's Claude family \cite{anthropic2023claude, anthropic2024claude}, we select claude-instant-1, claude-2, claude-3-haiku-20240307, and claude-3-5-sonnet-20240620  known for their code generation and problem-solving capabilities. Besides, we incorporate open-sourced deepseek-chat from DeepSeek \cite{guo2024deepseek} and qwen2.5-72b-instruct from Alibaba \cite{bai2023qwen}.

\noindent \textbf{Baselines. }
ReAct \cite{yao2022react} combines reasoning and action in a dynamic, step-by-step interaction, providing a flexible approach to task-solving. We use JSON as the action space. 
CodeAct \cite{wang2024codeact} utilizes a block of code as the LLM agent's action, enabling more efficient multi-turn interactions.

\noindent \textbf{Metrics. }
The evaluation includes accuracy and averaged turns. Accuracy represents the percentage of complex tasks that are correctly solved. 
We consider the LLM-generated code at the same layer, generated in parallel, as one turn.
We also record the average number of output words for the API cost evaluation.



\subsection{ToC vs. CodeAct and ReAct}
\begin{figure}[htbp!]
\centering
  \includegraphics[height=0.25\textwidth, width=0.48\textwidth]{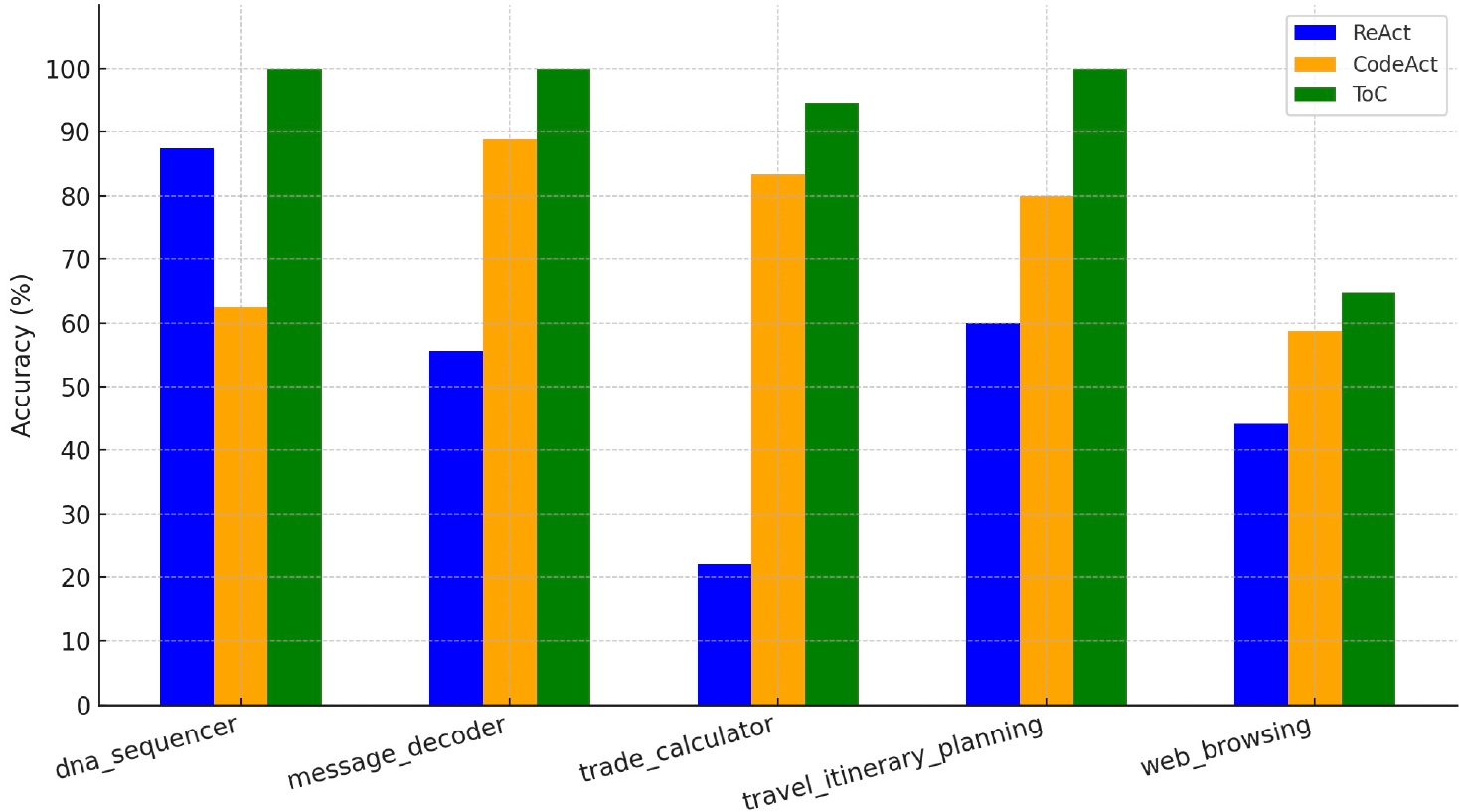}
  \caption{Comparison across five tasks in the M3.
  }
  \label{fig:chart}
  \vspace{-8pt}
\end{figure}

We primarily compare the ToC framework, which is comprised of CodeProgram nodes, with the CodeAct and ReAct framework, which are comprised of steps, using the M3 and the level-3 datasets.
For ToC, we randomly sample the LLM and prompt from the LLM list and prompt pool, respectively, at each node exploration. 
For CodeAct and ReAct, we report the average results across all LLMs used in this paper. 
Table \ref{tab:main} shows ToC achieves consistent superior performance (nearly 20\% higher) with significantly fewer interaction steps and averaged output words (nearly 1/4), highlighting its efficiency in handling complex tool-use scenarios.
Specifically, Figure \ref{fig:chart} shows the comparison of ReAct, CodeAct, and ToC on the five tasks in the M3, where ToC achieves near-perfect accuracy on all tasks except the web browsing task.

\subsubsection{Other multi-turn vs. Our one-turn}
Furthermore, we explore the performance of one-layer ToC (1-x) with the fixed model. 
As a node in ToC, CodeProgram enables the complete solution in a single turn by leveraging code’s ability to handle long logic chains.
Table \ref{tab:fix} shows that, with a significant advantage in the number of turns (one vs. multi-turn: averaged 7.0/8.9), our performance on some models even surpasses multi-turn CodeAct and ReAct, particularly with the Claude series.
Compared to CodeProgram, ie. ToC (1-1), the single layer, three nodes ToC (1-3) with random prompts significantly boosts performance. 
Its average accuracy already surpasses CodeAct, highlighting the effectiveness of prompt randomness.

We highlight the best-performing models in bold. Experimental results show that the top models differ between the CodeAct and ToC, and even within CodeAct, performance varies by dataset. For M3, gpt-4 performs best, while for API-Bank level-3, gpt-4o excels, likely because API-Bank level-3 emphasizes tool usage over scenario understanding, with simpler problem expressions. 
For ToC, claude-3-5-sonnet stands out due to its strong prompt-following ability, which is key for aligning reasoning with code and tool selection.

\subsection{Analysis and Ablation Studies}

\begin{table*}[htbp]
\centering
\renewcommand{\arraystretch}{1.35}
\scalebox{0.68}{
\begin{tabular}{l | c c c c | c c c c}
\toprule
\multirow{2}{*}{\textbf{Model}} & \multicolumn{4}{c|}{\textbf{M3ToolEval}} & \multicolumn{4}{c}{\textbf{API-Bank level-3}} \\ \cmidrule(lr){2-5} \cmidrule(lr){6-9}
 & \textbf{ReAct} & \textbf{CodeAct} & \textbf{ToC (1-1)} & \textbf{ToC (1-3)} 
 & \textbf{ReAct} & \textbf{CodeAct} & \textbf{ToC (1-1)} & \textbf{ToC (1-3)} \\
 
\midrule
\textbf{claude-instant-1} & 28.0\% (8.7) & 18.0\% (8.9) & 30.5\% (1) & 35.3\% (1) 
& 0.0\% (10.0) & 2.0\% (10.0) & 6.0\% (1) & 18.0\% (1) \\

\textbf{claude-2} & 40.2\% (8.2) & 54.9\% (7.2) 
& 57.3\% (1) & 59.8\% (1) 
& 0.0\% (10.0) & 20.0\% (8.9) & 8.0\% (1) & 18.0\% (1) \\

\textbf{claude-3-haiku} & 24.4\% (9.0) & 9.8\% (9.4) & 29.3\% (1) & 31.7\% (1) 
& 10.0\% (9.4) & 0.0\% (10.0) & 6.0\% (1) & 8.0\% (1)\\

\textbf{claude-3-5-sonnet} & 48.8\% (7.7) & 73.2\% (5.7)& \textbf{73.2\%} (1) & \textbf{82.9\%} (1) 
& 14.0\% (9.3) & 32.0\% (7.8) & \textbf{48.0\% }(1) & \textbf{52.0\% }(1) \\

\midrule
\textbf{gpt-3.5-turbo-1106} & 18.3\% (8.9) & 25.6\% (8.6) & 12.2\% (1) & 17.1\% (1) 
& 14.0\% (9.2) & 2.0\% (9.9) & 4.0\% (1) & 8.0\% (1) \\

\textbf{gpt-4-1106-preview} & \textbf{54.9\% }(7.5) & \textbf{75.6\% }(5.4) & 72.0\% (1) & 73.2\% (1) 
& \textbf{18.0\%} (8.2) & 30.0\% (8.2) 
& 34.0\% (1) & 38.0\% (1)\\

\textbf{gpt-4o-mini-2024-07-18} & 32.9\% (8.4) & 47.6\% (7.0) & 31.7\% (1) & 42.7\% (1) 
& 10.0\% (9.6) & 16.0\% (9.5) & 14.0\% (1) & 20.0\% (1)\\

\textbf{gpt-4o-2024-08-06} & 35.4\% (8.5) & 56.1\% (6.7) & 51.2\% (1) & 62.2\% (1) 
& 14.0\% (9.4) & \textbf{36.0\%} (7.8) & 28.0\% (1) & 32.0\% (1) \\

\midrule
\textbf{qwen2.5-72b-instruct} & 50.0\% (7.9) & 70.7\% (5.6) & 51.2\% (1) & 59.8\% (1) 
& 2.0\% (9.9) & 30.0\% (8.2) & 24.0\% (1) & 32.0\% (1)\\

\textbf{deepseek-chat} & 47.6\% (7.6) & 62.2\% (5.9) & 40.2\% (1) & 52.4\% (1) 
& 0.0\% (9.8) & 24.0\% (8.6) & 22.0\% (1) & 26.0\% (1)\\

\midrule
\textbf{\textit{Avg.}} & \textbf{38.05\% (8.24)} & \textbf{49.37\% (7.04)} & \textbf{43.53\% (1)} & \textbf{50.98\% (1) }
& \textbf{8.2\% (9.48)} & \textbf{19.2\% (8.89)} & \textbf{19.4\% (1)} & \textbf{24.4\% (1)}\\
\bottomrule
\end{tabular}
}
\caption{Ablation study of the model exploration. With different fixed models, the detailed performance comparison of ReAct, CodeAct, ablated ToC (1-1) (ie. the CodeProgram node), and ToC (1-3) on the M3ToolEval and API-Bank level-3 datasets is shown. The correctness is reported, with the average number of turns in parentheses.  
}
\label{tab:fix}
\end{table*}

\begin{table}[ht]
\centering
\renewcommand{\arraystretch}{1.1}
 \resizebox{\columnwidth}{!}{
\begin{tabular}{c|c|c|c}
\hline
\textbf{Layer / Node Per Layer} & \textbf{1}     & \textbf{2}     & \textbf{3}     \\ \hline
\textbf{1     }                         & 73.2\% (1)     & 75.6\% (1)     & 82.9\% (1)     \\ 
\textbf{2        }                      & 73.2\% (1.4)   & 76.8\% (1.4)   & 84.1\% (1.5)   \\
\textbf{3    }                          & 74.4\% (1.8)   & 79.3\% (1.7)   & 84.1\% (1.6)   \\ \hline
\end{tabular}
}
\caption{The performance of varying tree sizes.}
\label{tab:tree}
\end{table}
\noindent \textbf{Varying tree sizes.} 
We test the performance of the top model, claude-3-5-sonnet, on different tree sizes to evaluate the trade-off between efficacy and efficiency.
Table \ref{tab:tree} shows impressive results: 
with proper prompts and no additional training, 
the model achieves 84.1\% accuracy (3-3) on the M3, 10.9\% higher than 73.2\% (1-1). 

It seems the "nodes per Layer" contribute more than Layers, likely because our tree structure is designed to enhance exploration. Increasing the number of nodes certainly introduces more diverse prompts and model variations, whereas adding more layers (ie. more turns) mainly accumulates histories without significantly improving decision-making, especially with models that have limited contextual understanding. 

\noindent \textbf{Prompt exploration.} 
Ablation results in Table \ref{tab:ablation} confirm the effectiveness of prompt exploration. 
By comparing the random model with the fixed model (claude-3-5-sonnet), 
prompt exploration proves to be more critical in scenarios with lower diversity.

\begin{table}[ht]
\centering
\renewcommand{\arraystretch}{1} 
\resizebox{\columnwidth}{!}{
\begin{tabular}{l |c|c}
\toprule
\textbf{Mechanism} 
& \multicolumn{2}{c}{\textbf{M3ToolEval}} \\ \cmidrule(l){2-3}
                   & \textbf{Avg Turns} & \textbf{Correct} \\ 
\midrule
\multicolumn{3}{c}{Random Model ($\Delta=3.7$\%)} \\ \hline
ToC                & 1.7                & 67.1\%           \\ 
ToC w/o prompt exploration & 1.9        & 63.4\% $\downarrow$          \\ 
\midrule
\multicolumn{3}{c}{Fixed Model (the best) ($\Delta=8.5$\%)} \\ \hline
ToC w/o model exploration & 1.6         & 84.1\%           \\ 
ToC w/o model+prompt exploration & 1.8   & 75.6\% $\downarrow\downarrow$          \\ 
\bottomrule
\end{tabular}
}
\caption{Ablation study of the prompt exploration.}
  \label{tab:ablation}
\end{table}


\section{Related Work}
\textbf{LLM Code Generation for Complex Tasks.}
Recent works integrating LLMs with code have largely focused on task completion in programming domains like software development \cite{qian2024chatdev, wang2023leti}, programming assistance \cite{islam2024mapcoder, wen2024program}, and scientific problems \cite{chen2022pot, gao2023palsci, hong2024datasci}. These studies primarily address pure code generation, where correct task completion only relates to accurate reasoning logic within the code. For example, Chain of Codes \cite{li2023coc} broadens LLM capabilities by enabling "thinking in code."
In contrast, our work addresses real-world, zero-shot online complex tasks that involve multiple tool calls. Only CodeAct \cite{wang2024codeact} treats code as a scalable language to call multiple tools, but their approach is limited by an almost one-turn, one-tool, step-by-step mechanism. This results in stalled thinking and accumulated histories, relying heavily on ground-truth supervision for each step, which is incompatible with zero-shot, online settings. In our framework, every node represents a complete solution that can be directly evaluated via execution supervision without requiring additional labels.

\vspace{3pt}
\noindent \textbf{Tree-based Code Generation.}
A recent work, CodeTree \cite{Li2024CodeTreeAT}, uses a tree structure to explore the search space of code generation tasks. Unlike our approach, CodeTree focuses on multi-agent searching rather than an end-to-end, self-growing tree. 
While self-repair trees \cite{olausson2023self} begin with a specification root node, grow into initial programs through separate feedback and repair stages--often bottlenecked by the model’s limited capacity--our approach unifies reasoning (including reflection) and generation in a single cycle at each node, 
and directly expands the tree with prompt and model exploration.
Some contemporaneous works utilizing tree-based search, such as MCTS \cite{xu2024SRA-MCTS, yu2024outcome}, require multiple rollouts and significant computational resources, making them unsuitable for online, real-time applications. Unlike these methods, our self-growing tree generates multiple valid solutions and directly selects the one closest to the ground truth through a voting mechanism. Additionally, these studies typically focus on tasks with easier-to-obtain process supervision, whereas our work addresses real-world, complex multi-tool datasets.


\section{Conclusion}
This paper introduced the Tree-of-Code (ToC) method, which enables self-growing, end-to-end thought-code generation based on successful execution, addressing complex multi-tool online tasks. With efficient model integration and prompt exploration, ToC outperformed baselines on two complex task datasets, improving both efficiency and task-solving performance.




\section*{Limitations}

\subsection*{Limited reasoning scope for Program}
We emphasize that our method operates at the granularity of code "program" rather than "action". However, it is limited in fully open-ended scenarios requiring step-by-step exploration, such as a robot navigating an unfamiliar environment, or in handling tasks with extremely long sequences beyond the capabilities of current reasoning methods, like generating an entire paper. In such cases, it cannot provide a complete final solution. 
Even though, in practical industrial applications where a predefined toolset is available, CodeProgram’s end-to-end execution remains more efficient for online, zero-shot scenarios, for fewer turns, and for fewer LLM calls.

For larger and more complex system programs in the future, our method may serve as a "subprogram" within the overall solution, similar to a single agent's role in multi-agent systems.

\subsection*{Opportunities for Reflection Refinement}
While our framework provides a solid foundation inspired by human problem-solving, it uses a basic reflection mechanism, relying on execution feedback alone. Whether tracking full execution history or selectively summarizing with LLMs offers better performance remains an open question. Future research could explore enhanced search strategies or adaptive pruning methods to handle more complex real-world tasks.

\subsection*{Vast Potential in Prompt Pool Design}
We enhanced the diversity of strategies and the robustness of results in our Tree-of-Code by designing a prompt pool composed of multiple prompts. The introduction of multiple reasoning paths guided by diverse prompts represents a significant innovation. However, our current approach relies primarily on simple prompt evolution and manual adjustments. Future work should focus on more in-depth and systematic research into constructing prompt pools.

\subsection*{Acknowledgments}

This work was supported by the National Science and Technology Major Project under Grant 2022ZD0116409, and the National Natural Science Foundation of China under Grant 62301559.

\bibliography{ToC}

\newpage
\appendix
\label{section:appendix}
\onecolumn

\section{Prompt}
\label{Appendix Prompt}

\subsection{Root Prompt}
\label{Root Prompt}
\begin{lstlisting}[breaklines=true, basicstyle=\scriptsize]
You are a helpful assistant assigned with the task of problem-solving. 
To achieve this, you will be using an interactive coding environment equipped with a variety of tool functions to assist you throughout the process.\n\n
At each turn, you should first provide your step-by-step thinking for solving the task, for example: <thought> I need to print "hello world!"</thought>. 
After that, you can Interact with a Python programming environment and receive the corresponding output. 
Your code should be enclosed using "<execute>" tag, for example: <execute> print("Hello World!") </execute>.\n\n 
You can use the following functions:\n{toolset_descs}\n. 
Ensure the code matches the fn_signature and input-output formats for proper execution.\n
Here's the chat history for your reference:\n{chat_history}\n\n
History End:\n
User's Query:\n{query}\nYour Thought And Code:\n
\end{lstlisting}

\subsection{Additional Prompt}
\subsubsection{Reflection Prompt}
\label{Reflection Prompt}
\begin{lstlisting}[breaklines=true, basicstyle=\scriptsize]
Based on the provided chat history, reflect on the code and its execution. Identify potential issues or areas for optimization and provide specific suggestions to refine and improve the code. Consider edge cases, efficiency, and clarity in your reflections.
\end{lstlisting}

\subsubsection{The Prompt for Prompt Evolution}
\label{Prompt Evolution}
\begin{lstlisting}[breaklines=true, basicstyle=\scriptsize]
In order to guide the diversity of results and enhance the performance through ensemble methods, we need to increase the diversity of prompts. We diversify the current prompt while maintaining consistency in core content, aiming for orthogonal expressions or prompts that lead to different directions and divergent thinking.
\end{lstlisting}

\subsubsection{The Prompt Sample from Prompt Pool for API-Bank}
\label{Helper Prompt Sample for API-Bank}
\begin{lstlisting}[breaklines=true, basicstyle=\scriptsize]
Note:
The outputs produced by the tool will be formatted like a JSON dictionary. 
For example, 'result = {{'api_name': 'QueryMeeting', 'input': {{'user_name': 'John'}}, 'output': {{'meetings': [{{'meeting_id': 1, 'meeting_name': 'Meeting with the client', 'meeting_time': '2021-01-01 10:00:00', 'meeting_location': 'Room 1', 'meeting_attendees': ['John', 'Mary', 'Peter']}}, {{'meeting_id': 2, 'meeting_name': 'Meeting about the new project', 'meeting_time': '2021-01-02 10:00:00', 'meeting_location': 'Room 2', 'meeting_attendees': ['John', 'Mary', 'Peter']}}]}}, 'exception': None}}'
Ensure that the code strictly adheres to the function descriptions and the input-output format provided.
Navigate through the 'output' key correctly to retrieve results.
If you encounter any unfamiliar formats, first print the structure to ensure proper handling in the future.
Consistently focus on the user's request and attempt to produce the complete solution without needing multiple steps.
\end{lstlisting}

\section{Helper tools}
\label{Helper Tool}

\subsection{ResHandler}
\subsubsection{ResHandler Tool Description}

\begin{lstlisting}[breaklines=true, basicstyle=\scriptsize]
res_handler(): 
    name="res_handler",
    description='Define a prompt to generate results that meet the prompt requirements. Note that you need to define the requirements for the generated results in the prompt. input: prompt (str): The input prompt for the large language model, defining the task requirements for the generated results. Common tasks include summarization, stylistic writing, translation, question answering, etc. output: completion (str): The inference result generated by the large model, typically a summary, writing output, translation result, or answer that meets the requirements.',
    function=res_handler,
    fn_signature='res_handler(prompt: str) -> str')
\end{lstlisting}

\subsubsection{ResHandler Tool Function}

\begin{lstlisting}[breaklines=true, basicstyle=\scriptsize]
from some_model_API import llm_playground

def res_handler(prompt):
    result_str = ""
    result = llm_playground(prompt[:20000], stream=False)
    for item in result:
        result_str += item
    return result_str

\end{lstlisting}

\subsection{NextAction for Web Task}
\label{NextAction}
\subsubsection{NextAction Tool Description}

\begin{lstlisting}[breaklines=true, basicstyle=\scriptsize]
from typing import Tuple
next_action():
    name="next_action", 
    description='Examine the results of the view function to determine if it can answer the user's original question, and decide what to do next. Return the next action and the viewed whole page content.The next possible actions include click_url(URL), go_to_previous_page() and end(), which represent clicking a link, and go_to_previous_page() means you should go to previous page to find answer, and end() means you have found the answer page, respectively. If next action is end(), it means that relevant information to user query is found, you should summarize string result based on res_handler. click_url(URL), go_to_previous_page() can be directly called, and URL should be Clickable url. Note that query should be user's original question and can not be rewritten.',
    function=next_action,
    fn_signature="next_action(query: str, current_page_content: str, visited_urls: List[str]) -> Tuple[str, str]")
\end{lstlisting}

\subsubsection{NextAction Tool Description}
\begin{lstlisting}[breaklines=true, basicstyle=\scriptsize]
from some_model_API import llm_playground

def next_action(query="", current_page_content="", visited_urls=[]):
    visited_urls = [x.replace('\'', '').replace('\"', '') for x in visited_urls]
    visited_urls = list(set(visited_urls))
    whole_page_content = current_page_content
    while True:
        scroll_down_page = scroll_down()
        if scroll_down_page == "[Reached the bottom of the page.]\n":
            break
        else:
            whole_page_content += scroll_down_page
    def extract_clickable_paths(text: str) -> list[str]:
        import re
        pattern = r"Clickable '([^']*)'"
        matches = re.findall(pattern, text)
        return matches
    all_urls = extract_clickable_paths(whole_page_content)

    not_visited = []
    highlight_urls = []
    
    for v in all_urls:
        if v in visited_urls:
            highlight_urls.append(v)
        else:
            not_visited.append(v)

    if len(highlight_urls) == 0:
        json_str_format = "<thought>your thought of your decision</thought>\n<action>click_url(specific_url) or end() or not_found()</action>"
        prompt = f"You are viewing page contents, the content is: \n{whole_page_content}\n You should make decision on the next step. given user query {query}, you have the following options, please follow the output format. \n1. end(): it means current user query can be answered by current page content. \n2. click_url(URL): it means current user query should be checked by clicking one of the urls shown on the current page content for more details. specify the detailed url into URL field.\nPlease visit any Clickable urls as many as possible that has not been visited. \n3. not_found(): it means that current page does not contain answer for current query and all Clickable URLS have been clicked. \nYour output format: {json_str_format}\n\nYour Output:\n"
    else:
        visited_url_str = ', '.join(['\'' + x + '\'' for x in highlight_urls])
        json_str_format = "<thought>your thought of your decision</thought>\n<action>click_url(specific_url) or end() or not_found()</action>"
        prompt = f"You are viewing page contents, the content is: \n{whole_page_content}\n You should make decision on the next step. given user query {query}, you have the following options, please follow the output format. \n1. end(): it means current user query can be answered by current page content. \n2. click_url(URL): it means current user query should be checked by clicking one of the urls shown on the current page content for more details. specify the detailed url into URL field.\n3. not_found(): it means that current page does not contain answer for current query and all Clickable URLS have been clicked. \nRemember that you have visited the url list [{visited_url_str}]. You are not allowed to visit the urls you have visited. Please visit any Clickable urls as many as possible that has not been visited.\nYour output format: {json_str_format}\n\nYour Output:\n"
    result_str = ""
    result = llm_playground(prompt[:20000])
    for item in result:
        result_str += item

    if not "Clickable" in whole_page_content and not "end()" in result_str:
        return ("go_to_previous_page()", whole_page_content)

    if not "end()" in result_str and len(not_visited) == 0:
        return ("go_to_previous_page()", whole_page_content)
        
    if "click_url" in result_str:
        import re
        pattern = r"click_url\('.*'\)"
        match = re.search(pattern, result_str)
        if match:
            return (match.group(), whole_page_content)
        else:
            pattern = r"click_url\(.*\)"
            match = re.search(pattern, result_str)
            if match:
                return (match.group(), whole_page_content)
    elif "end()" in result_str:
        return ("end()", whole_page_content)
    elif "not_found()" in result_str:
        return ("go_to_previous_page()", whole_page_content)        
    return ("end()", whole_page_content)
\end{lstlisting}
   
\subsection{An example to show the performance of two helpers tools}
\label{Helper Tool Example}
For web-based queries, ResHandler and NextAction, these two helper tools are essential to ensure that our method can directly generate the final result without requiring intermediate feedback.

To further clarify their effectiveness, we provide a qualitative and specific example demonstrating how they function in practice. 

Below is the full code output for the query: "Find the email of Henry Santiago. Answer in the format of 'xxx@xxx.xxx'." 
from typing import List, Tuple
\begin{lstlisting}[breaklines=true, basicstyle=\scriptsize]
def find_email(query: str, current_page_content: str, visited_urls: List[str], max_depth: int = 30) -> str:
    if max_depth <= 0:
        raise Exception("Max search depth reached, email not found.")
    
    # Check if the email is directly available in the current page content
    if "Henry Santiago" in current_page_content and "@" in current_page_content:
        email = res_handler(f"Extract the email of Henry Santiago from the following text: {current_page_content}")
        if email:
            return email.strip()
    
    # Identify clickable URLs and navigate to them
    next_step_action, reference_page_content = next_action(query, current_page_content, visited_urls)
    
    if next_step_action == "end()":
        email = res_handler(f"Extract the email of Henry Santiago from the following text: {reference_page_content}")
        if email:
            return email.strip()
        else:
            raise Exception("Email not found in the reference page content.")
    
    elif next_step_action.startswith("click_url"):
        url = next_step_action.split('(')[1].split(')')[0].strip().strip('"')
        if url not in visited_urls:
            visited_urls.append(url)
            new_page_content = click_url(url)
            return find_email(query, new_page_content, visited_urls, max_depth - 1)
    
    elif next_step_action == "go_to_previous_page()":
        new_page_content = go_to_previous_page()
        return find_email(query, new_page_content, visited_urls, max_depth - 1)
    
    else:
        raise Exception("Unknown next step action encountered.")

# Main function to start the search
def main():
    query = "Find the email of Henry Santiago"
    initial_page_content = view()
    visited_urls = []
    try:
        email = find_email(query, initial_page_content, visited_urls)
        print(email)
    except Exception as e:
        print(f"Error: {e}")

if __name__ == "__main__":
    main()


\end{lstlisting}

\noindent \textbf{Functionality of the Helper Tools}

\noindent For res\_handler: 

Example usage:
\begin{lstlisting}[breaklines=true, basicstyle=\scriptsize]
    email = res_handler("Extract the email of Henry Santiago from the following text: {current_page_content}")
\end{lstlisting}

We provide the content that the final result depends on (i.e., the return value of other functions in the code) as input to this function.
It then passes the information to an LLM to generate the final answer.

\noindent For next\_action:

Example usage:
\begin{lstlisting}[breaklines=true, basicstyle=\scriptsize]
    # Identify clickable URLs and navigate to them 
    next_step_action, reference_page_content = next_action(query, current_page_content, visited_urls)
\end{lstlisting}

This function determines the next action based on: the current page content, the original query, and the list of visited URLs.

\section{Visualization of the Table 2}
\begin{figure}[H]
\centering
  \includegraphics[width=1\textwidth]{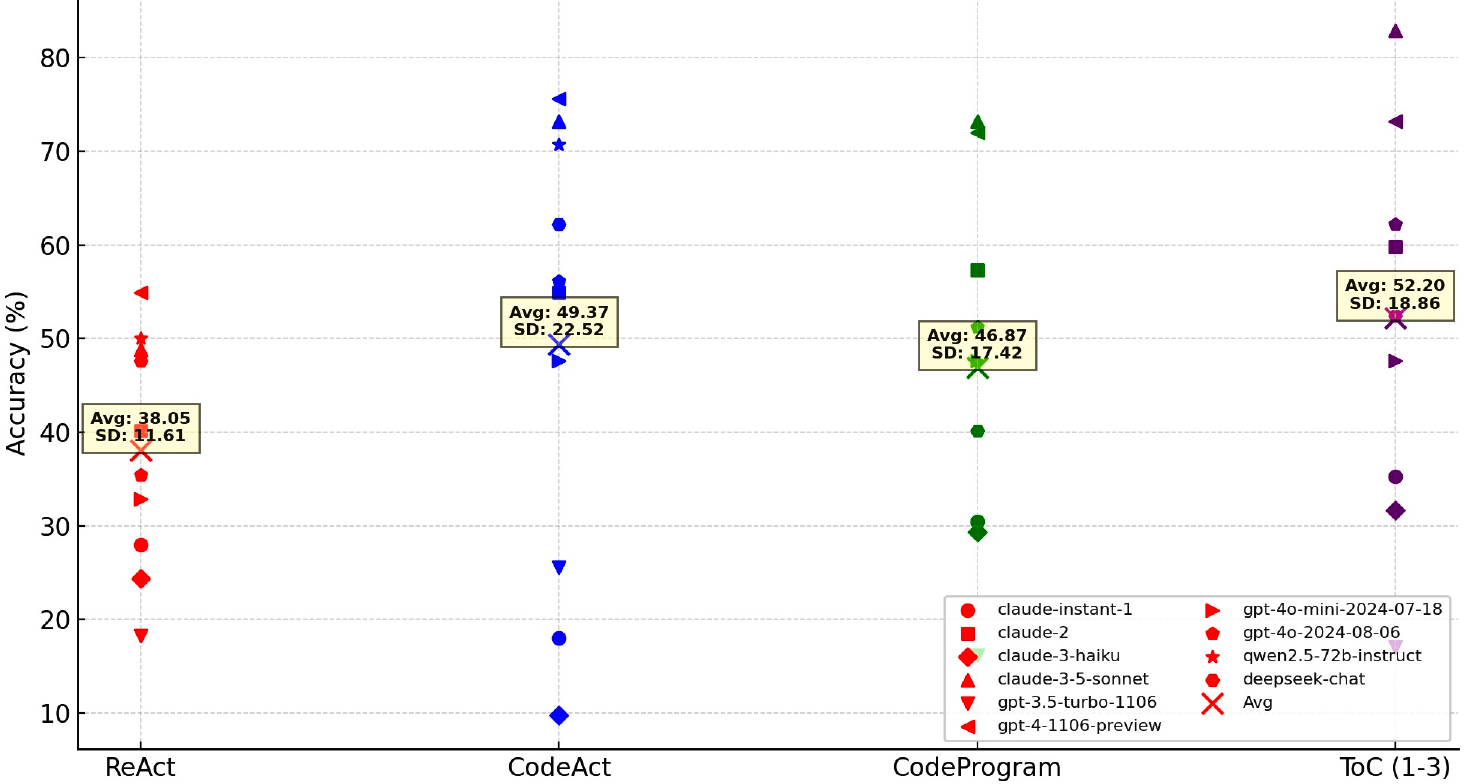}
  \caption{
  Performance of 10 LLMs on ReAct, CodeAct, CodeProgram, and 1-3 ToC for the M3 dataset is visualized, with average and standard deviation reported.
  }
  \label{fig:std}
\end{figure}

\end{document}